# Grover's algorithm based multi-qubit secret sharing scheme

**Arti Chamoli and C. M. Bhandari**
**Indian Institute of Information Technology, Allahabad, Deoghat, Jhalwa, Allahabad-211011, India.**
**Email: achamoli@iiita.ac.in, cmbhandari@yahoo.com**

**Abstract**
Some of the secret sharing schemes having unique quantum features like parallelism and entanglement are supposed to be relatively secure. Different schemes proposed by various researchers over the years have features which could be specific to the nature and need of a situation. Following Hsu's proposed scheme we propose a secret sharing scheme using Grover's search algorithm for a four qubit system with several marked states. Further, the scheme has been generalized to an n-qubit system.

## Introduction

Quantum secret sharing can be accredited as one of the major implications of quantum information processing. It addresses the problem of secured transfer of information through quantum channels to distant receivers. With the advent of quantum computers, quantum secret sharing schemes are being extensively studied by several researchers. These quantum counterparts of the classical secret sharing schemes exploit unique quantum features like parallelism and entanglement. The message to be communicated is encrypted in arbitrary quantum states and is sent to intended receivers after initial quantum manipulations by the sender. These manipulations are so performed that none of the receivers can retrieve the complete secret message single-handedly. The idea of quantum secret sharing scheme was originally conceived and implemented by Hillery et al [1] using three and four particle Greenberger-Horne-Zeilinger (GHZ) states [2]. Similar to the protocol given in [1], Karlson et al [3] showed the implementation of a quantum secret sharing protocol using two-particle quantum entanglement. Later, (k,n)-threshold quantum secret sharing scheme was proposed by Cleve et al [4] in which the encoded quantum state is split among $n$ people such that any $k$ of them can reconstruct the encoded information, while number of people, if less than $k$, can never succeed. This has been experimentally demonstrated in [5,6]. Gottesman [7] demonstrated mixed state quantum secret sharing by discarding a share from pure state scheme, the only constraints being monotonicity [8] and the no cloning theorem [9-11]. This quantum secret sharing scheme for general access structures was shown by Smith [12] in a somewhat different manner. Karimpour et al [13] demonstrated quantum secret sharing schemes based on entanglement swapping of generalized d-level Bell states. Cabello [14] generalized quantum secret sharing schemes to n-particles. Recently, Guo-Ping Guo et al [15] presented quantum secret sharing without entanglement.

In 2003, a new dimension was added to quantum secret sharing schemes with the incorporation of Grover's unsorted database search algorithm [16]. Li-Yi-Hsu proposed a two-qubit quantum secret sharing protocol based on Grover's algorithm [17]. Apart from the operational aspects the protocol is different from the above mentioned schemes in the sense of cheat detection. With this protocol cheating could be detected immediately without exhausting a portion of the sequence of measurements of outcomes. In a two-qubit Grover's algorithm, the marked state can be retrieved with full probability after a single iteration. Hsu's protocol is based on this property. In this paper we present a generalization of two-qubit quantum secret sharing protocol based on Grover's algorithm to n-qubit system. We will show that marked state can be found with certainty after a single iteration for any number of qubits in Grover's search algorithm if the number of marked states is one-fourth of the total number of elements in search space. Thus two-qubit quantum secret sharing protocol based on Grover's search algorithm can be generalized to n-qubit secret sharing scheme. The increase in the number of marked states with the increase in number of qubits not only reduces the probability of error but also enhances the security aspect of the secret sharing protocol. The message can be split and encoded into various marked states. We start with a review of original scheme presented by Hsu.

**Two qubits secret sharing protocol**

Alice, the sender, randomly prepares a two-qubit superposition state $|S_1\rangle$ of the form

$$\left(1/\sqrt{2}\right)\left(|0\rangle+|1\rangle\right)\otimes\left(1/\sqrt{2}\right)\left(|0\rangle+|1\rangle\right)$$

The superposition state can be the product of any two of the following four states: $\left(1/\sqrt{2}\right)\left(|0\rangle+|1\rangle\right)$, $\left(1/\sqrt{2}\right)\left(|0\rangle-|1\rangle\right)$, $\left(1/\sqrt{2}\right)\left(|0\rangle+i|1\rangle\right)$, $\left(1/\sqrt{2}\right)\left(|0\rangle-i|1\rangle\right)$. The message is encoded in the marked states $|01\rangle$ or $|10\rangle$ , whereas the states $|00\rangle$ or $|11\rangle$ are used to detect any possible eavesdropping.

She performs $P_W$ operation on $|S_1\rangle$ where $P_W$ is of the form

$$P_W = 1-2|W\rangle\langle W|, \tag{1}$$

$|W\rangle$ being the marked state contains the secret information. If the message is encrypted in the state $|10\rangle$, then

$$\begin{aligned} P_W\,|S_1\rangle = |S_1\rangle_W &= \left[1-2|10\rangle\langle 10|\right]\left[(1/2)\left(|00\rangle+|01\rangle+|10\rangle+|11\rangle\right)\right] \\ &= \left[(1/2)\left(|00\rangle+|01\rangle-|10\rangle+|11\rangle\right)\right] \end{aligned} \tag{2}$$

This transformation changes the phase of the desired state keeping the other states unchanged. Alice then sends these two qubits to Bob and Charlie who are at a distant place. The qubits sent by Alice are such that Bob receives the first qubit and Charlie the second one. It is assumed that at most one of them may be a cheat who may try to capture both qubits so that he can retrieve whole of the information by himself. Hence Alice also aims to detect any possible cheating strategy.

Once she has confirmed via classical channel that each one of them is in receipt of their respective qubit, she makes a public announcement of her initial state $|S_1\rangle$ . Even after Alice's declaration of the initial state $|S_1\rangle$ Bob and Charlie can have access to the encrypted information only when they combine their respective qubits and perform $-P_{S_1}$ on the two qubits collectively.

$$
\begin{aligned}
&-P_{S_1}|S_1\rangle_W = \left[2|S_1\rangle\langle S_1| - I\right]\left[(1/2)(|00\rangle + |01\rangle - |10\rangle + |11\rangle)\right] \\
&= \left[2\{(1/2)(|00\rangle+|01\rangle+|10\rangle+|11\rangle)\}\{(1/2)(\langle 00|+\langle 01|+\langle 10|+\langle 11|)\} - I\right]\left[(1/2)(|00\rangle+|01\rangle-|10\rangle+|11\rangle)\right] \\
&= |10\rangle
\end{aligned} \tag{3}
$$

Thus the secret information could only be gained collectively.

Alice can check any possible eavesdropping by encrypting her secret information only in the states $|01\rangle$ or $|10\rangle$ while she uses states $|00\rangle$ and $|11\rangle$ for cheat detection. Hence, according to the protocol, Bob and Charlie are supposed to inform Alice if their outcomes of collective measurements are correlated. If the result of measurement comes out to be correlated then honest receiver concludes that eavesdropping has occurred or Alice has encrypted nothing. In either case, the receivers inform Alice about correlated outcomes. The process is aborted if Alice detects any eavesdropping otherwise she continue sending message to the intended receivers.

**n-qubit Grover's algorithm after one iteration**.

The Grover's algorithm addresses the following problem: for a search space of $N$ items with $M$ marked states within it such that $1 \le M \le N$ , Grover's algorithm can find the marked state in $O\left(\sqrt{N/M}\right)$ iterations. We have shown below the state of the n-qubit system after iterating the algorithm once. For simplicity and without loss of generality we can assume that $N = 2^n$ elements.

For a search space $S$, of $N = 2^n$ elements, we can represent the $N$ elements using an n-qubit register with their indices $i = 0, -----N-1$. If a set of $M$ elements within the search space represents the marked states, that is, they are solution to the search problem, then the marked states can be identified by a function $f$ which maps the elements of $S$ to either 0 or 1, $f: S \to \{0,1\}$, such that $f = 1$ for the marked elements only. The steps of the Grover's algorithm are as follows:

1. *Register preparation* $\to$ A register $|i\rangle^{\otimes n}$ of n qubits and an ancilla qubit $|0\rangle$ to evaluate the oracle is prepared.

$$
|X_1\rangle = |0\rangle^{\otimes n} \otimes |0\rangle \tag{4}
$$

2. *Initialization*$\rightarrow$ Apply *H*, Hadamard gate, to each of the n qubits in the register, and the *HX* gate to the ancilla qubit, where *X* is a NOT gate.

$$\begin{aligned}|X_1\rangle &= \left(H^{\otimes n}\otimes I\right)|0\rangle^{\otimes n} HX\,|0\rangle \\ &= \frac{1}{\sqrt{N}}\sum_{i=0}^{N-1}|i\rangle\otimes\left(\frac{|0\rangle-|1\rangle}{\sqrt{2}}\right)\end{aligned} \quad (5)$$

3. *Applying the oracle*$\rightarrow$ The oracle is applied in the form of operator $P_W$ which flips the state of ancilla qubit on identifying the marked state.

$$P_W\,|i\rangle|q\rangle = |i\rangle|q\oplus f(i)\rangle$$

If $|i\rangle$ is a marked state, then $f(i)=1$. Since the state of ancilla qubit is $\left(\frac{|0\rangle-|1\rangle}{\sqrt{2}}\right)$, the action of $P_W$ can also be written as

$$P_W\,|i\rangle\left(\frac{|0\rangle-|1\rangle}{\sqrt{2}}\right) = (-1)^{f(i)}\,|i\rangle\left(\frac{|0\rangle-|1\rangle}{\sqrt{2}}\right) \quad (6)$$

As the state of ancilla remains unchanged, the above expression can be rewritten as

$$P_W\,|i\rangle = (-1)^{f(i)}\,|i\rangle$$

Thus the effect of the operator $P_W$ on the superposed n-qubit register can safely be expressed as

$$|X_2\rangle = P_W\,\frac{1}{\sqrt{N}}\sum_{i=0}^{N-1}|i\rangle = \frac{1}{\sqrt{N}}\left[\sum_{i=0}^{N-1}|i\rangle\right]_N - \frac{1}{\sqrt{N}}\left[\sum_{i=0}^{N-1}|i\rangle\right]_M \quad (7)$$

For M marked states, $\left[\sum_{i=0}^{N-1}|i\rangle\right]_M$ indicates the summation over all *i* which are the marked states, and $\left[\sum_{i=0}^{N-1}|i\rangle\right]_N$ indicates the summation over all *i* which are the remaining elements of the search space.

3. *Applying inversion about average operator* $\rightarrow$ The application of operator $P_s$, which performs inversion about average results into

$$|X_3\rangle = P_S|X_2\rangle = \left\{2\frac{1}{\sqrt{N}}\sum_{i=0}^{N-1}|i\rangle\frac{1}{\sqrt{N}}\sum_{i=0}^{N-1}\langle i| - I\right\}\left\{\frac{1}{\sqrt{N}}\left[\sum_{i=0}^{N-1}|i\rangle\right]_N - \frac{1}{\sqrt{N}}\left[\sum_{i=0}^{N-1}|i\rangle\right]_M\right\}$$

$$|X_3\rangle = \left\{\frac{N-4M}{N}\right\}\frac{1}{\sqrt{N}}\left[\sum_{i=0}^{N-1}|i\rangle\right]_N + \left\{\frac{3N-4M}{N}\right\}\frac{1}{\sqrt{N}}\left[\sum_{i=0}^{N-1}|i\rangle\right]_M$$

(8)

This represents the state of the system after iterating the algorithm once. It can be easily verified that if a measurement is made the coefficient of the unmarked states becomes zero for $M = N/4$, that is, the number of marked states is one-fourth of the total search space. Thus one of the marked states can be found with certainty if a measurement is made after a single iteration.

The probability $P_S$, of finding one out of M marked states after first iteration can be calculated from eq.(8),

$$P_S = M\left(\frac{3N-4M}{N\sqrt{N}}\right)^2$$

$$= 9\left(\frac{M}{N}\right) - 24\left(\frac{M}{N}\right)^2 + 16\left(\frac{M}{N}\right)^3,$$

Similarly, the probability $P_S'$, of finding an undesired item is

$$P_S' = (N-M)\left(\frac{N-4M}{N\sqrt{N}}\right)^2.$$

$$\text{Obviously, } P_S + P_S' = 1.$$

(9)

Fig.1. shows the plot of success probability after single iteration for $0 < M/N \le 1$. Success probability becomes one, that is one of the marked states can be found with certainty only when $M/N = 1/4$.

**Four-qubit secret sharing scheme**

The initial superposition state of four qubits can be a product of any of the following single qubit states: $(1/\sqrt{2})(|0\rangle + |1\rangle)$, $(1/\sqrt{2})(|0\rangle - |1\rangle)$, $(1/\sqrt{2})(|0\rangle + i|1\rangle)$, and $(1/\sqrt{2})(|0\rangle - i|1\rangle)$ denoted as $|+\rangle, |-\rangle, |+i\rangle, |-i\rangle$. Suppose Alice, the sender, prepares the superposition state $|S_1\rangle$ of the form,

$$|S_1\rangle = \left[\left(\frac{1}{\sqrt{2}}\right)(|0\rangle + |1\rangle)\right]^{\otimes 4}$$

$$= \frac{1}{4}(|0000\rangle + |0001\rangle + |0010\rangle + |0011\rangle + |0100\rangle + |0101\rangle + |0110\rangle + |0111\rangle + |1000\rangle + |1001\rangle + |1010\rangle + |1011\rangle + |1100\rangle + |1101\rangle + |1110\rangle + |1111\rangle) \quad (10)$$

She then encodes the message in the marked states. As discussed earlier, the number of marked states in this case will be four. She encodes half of the message in any two of the four marked states and the remaining message in the other two marked states. Same message can be encoded in different bit strings, each of which represents the element of the search space. This is because the oracle recognizes and flips the sign of the marked state by considering a function $f$ which maps the marked state, with specific queries, to 1. The function $f$ is so designed that it identifies two elements of the search space for same query. This can be interpreted as tagging the same message to two different bit strings.

After encrypting the secret message, Alice applies the operator $P_W$ on the initial state $|S_1\rangle$, which inverts the phase of the marked states. Suppose the marked states are $|0100\rangle$, $|0110\rangle$, $|1000\rangle$ and $|1011\rangle$ then the state of the system evolves to an entangled state

$$P_W|S_1\rangle = \frac{1}{4}(|0000\rangle + |0001\rangle + |0010\rangle + |0011\rangle - |0100\rangle + |0101\rangle - |0110\rangle + |0111\rangle - |1000\rangle + |1001\rangle + |1010\rangle - |1011\rangle + |1100\rangle + |1101\rangle + |1110\rangle + |1111\rangle) \quad (11)$$

Alice then sends all four qubits to the distant receivers to whom the message is to be communicated. The receivers are informed in advance about the order of the bit strings in the sense that who shall receive the first qubit and so on. Here it can be said that even if three out of four receivers are dishonest, they cannot decipher the secret information without the joint effort of all four receivers.

At this stage, Alice confirms classically that each of the receivers have received their respective qubits. It is assumed that the dishonest receiver(s) or for that matter any eavesdropper may try to get access to the secret information without any assistance from Charlie. Hence the receivers are supposed to inform her about receiving their respective qubits. Since she does not know who the cheat is therefore she announces her initial preparation $|S_1\rangle$ only after getting confirmation from all four. Now the honest receiver(s) combine his qubit with other qubits and perform $-P_{S_1}$ to retrieve the secret message. The dishonest member(s) cannot do any mischief while performing the collective operation as the honest member(s) can check him.

After performing the inversion about average operation, $-P_{S_1}$, the receivers will obtain any one of the four marked states.

$$-P_{S_1}\ P_W\,|S_1\rangle = \frac{1}{2}\left[|0100\rangle + |0110\rangle + |1000\rangle + |1011\rangle\right]$$

As the quantum searching process is probabilistic, Alice needs to send two or more sets of $P_W\,|S_1\rangle$ in order to communicate the complete message. The probability of error at the receiving end is reduced because each half of the message occurs twice with same probability.

One of the possible ways of cheat detection is that Alice does not encrypt any secret information and sends the state $P_W\,|S_1\rangle$ just to check the authenticity of the intended receivers. She then asks the receivers to inform her classically about their measurement outcome. If the outcome differs from the four marked states, she immediately detects cheating with a probability of $11/16$ and stops the process. The dishonest receiver(s) cannot cheat by giving false information about his outcome because he does not know what the marked state is. The dishonest receiver(s) or any eavesdropper may try to capture all the four qubits therefore the sender and the honest receiver(s) must check any possible eavesdropping.

Suppose an eavesdropper or the dishonest receiver(s) succeeds in getting hold of all the four qubits. His aim will be to retrieve the secret information all by himself. The number of possible four-qubit product states with $|+\rangle, |-\rangle, |+i\rangle, |-i\rangle$ would be 256. The possibility of dishonest member getting access to the secret message with certainty is only if he performs the correct $-P_{S_1}$ operation. The probability of which is $1/256$. If he applies the wrong $-P_{S_1}$ operator, the probability of getting the correct state will be very low while Alice's probability of cheat detection will increase because the number of unmarked states is quite large than the marked states.

Secondly, the cheater may capture the state sent by Alice and send some other state to the legitimate receivers. Thus after Alice's announcement of the initial state, he can transform the captured state accordingly. The probability of cheat detection is very high since he does not know the state $P_W\,|S_1\rangle$. For a general four qubit product state $|S_i\rangle$ with four marked states there are 1820 possible $P_W\,|S_i\rangle$. The probability of sending the correct $P_W\,|S_i\rangle$ and going undetected is $1/1820$. He may send the state $P_W\,|S_i\rangle$ in which out of the four chosen marked states only one or two or three marked states tally with Alice's marked states. Thus the total probability of not getting detected is $1/728$.

Table1. shows the outcomes of some of the $-P_{s_i}\ P_W\,|S_1\rangle$, where $i = 1.....10$ represent the possible initial states. Only on applying the correct $-P_{s_i}$ operator, the measurement

outcome will be one of the marked states with certainty. In any other case the probability of getting one of the marked states is remarkably reduced.

This four-qubit secret sharing scheme based on Grover's algorithm can be generalized to n-qubit secret sharing scheme. For an n-qubit secret sharing scheme, there are $n$ legitimate users. Alice, the quantum information sender, prepares the initial n-qubit product state. The initial state of this quantum secret sharing scheme should be a product state because the initial state of Grover's search algorithm is necessarily a product state for maximum success probability.

Alice prepares a n-qubit initial product state having $N$ substates. She encodes the secret information in M of the N substates. These M states constitute the marked states. As discussed earlier M should be one-fourth of N. The message is split into two halves. The encryption of the secret message is such that $M/2$ substates are tagged with half of the message and the other $M/2$ substates with remaining part of the message. Alice then applies the operator $P_W$ on the initial state. W denotes the M marked states. She then sends all the n qubits to respective receivers. After confirming that all receivers are in receipt of their respective qubits, she announces her initial $|S_i\rangle$ in public. The receivers can perform the operation $-P_{S_1}$ when they combine their qubits. The application of $-P_{S_1}$ gives one of the marked states with certainty. Since quantum measurements are probabilistic therefore Alice needs to send few more sets of $P_W|S_i\rangle$ for complete transfer of information. The advantage of labeling $M/2$ states with same portion of the message is to increase the probability of obtaining the message with higher accuracy, that is, the error is minimized. Also the splitting of message into two halves decreases the success probability of eavesdropper. The detection probability increases with the number of qubits because the eavesdropper cannot determine the initial state or the exact combination of marked states with certainty.

**Single marked state**

Another quantum secret sharing protocol based on Grover's algorithm with single marked state is considered. Hsu's protocol for a two qubit system with single marked state could be extended to three, four, and five-qubit entangled states. As the number of qubits increases to three or four, the probability of obtaining the desired state is somewhat decreased hence the sender may be required to send four to five identical sets of qubits to the distant receivers. However, for a five qubit system, the probability of getting the desired state with certainty becomes similar to that of a two qubit system. The number of iterations in this case would be more. For example, for a five qubit system total number of iterations will be four and the sender will have to transfer the qubits after performing three complete iterations and operation $P_W$ of the fourth iteration. The number of receivers will be of the order of number of qubits so that cheating becomes relatively difficult. Also, the probability of applying the correct operator by any cheat decreases with the increase in initial superposition states.

We now exemplify our analysis for a three qubit system. Alice wants to send some secret information to a distant place where she has three agents who are supposed to receive it. She follows the following steps.

1. She initializes the system to a superposition of $2^3 = 8$ states by performing some local operations. The amplitude of each of the eight states is $\left(1/2\sqrt{2}\right)$. She randomly prepares some state $|\psi\rangle$ (say)

$$|\psi\rangle = \left(1/2\sqrt{2}\right)\left(|000\rangle+|001\rangle+|010\rangle+|011\rangle+|100\rangle+|101\rangle+|110\rangle+|001\rangle\right)$$

2. After preparing state $|\psi\rangle$ she performs unitary operation $P_W$ on it. Let $W$ be $|110\rangle$ for the present case.

$$P_W\,|\psi\rangle = |\psi\rangle_W = \left(1/2\sqrt{2}\right)\left(|000\rangle+|001\rangle+|010\rangle+|011\rangle+|100\rangle+|101\rangle-|110\rangle+|111\rangle\right)$$

This changes the phase of the desired state. Now she performs $-P_\psi$ on $|\psi\rangle_W$ and gets

$$\left(1/4\sqrt{2}\right)\left(|000\rangle+|001\rangle+|010\rangle+|011\rangle+|100\rangle+|101\rangle\right)+\left(5/4\sqrt{2}\right)\left(|110\rangle\right)+\left(1/4\sqrt{2}\right)\left(|111\rangle\right)$$

This increases the amplitude of the desired states thereby decreasing the amplitude of rest of the states. This is her first iteration.

3. For the second iteration she performs only $P_W$ and then sends her three qubits to her three distant agents Bob, Charlie and Trent who are supposed to decrypt the secret message only if they work collectively.

At this stage she confirms classically that each of them received their respective qubits. She announces her initial preparation only after getting confirmation from all three. Now the three receivers combine their qubits and perform $-P_\psi$ to retrieve the secret message. This leads to the completion of second iteration. The dishonest members cannot do any mischief while performing the collective operation as the honest member can check them.

4. Thus the amplitude of the desired state rises to $(11/8\sqrt{2})$ thereby decreasing the amplitude of other states to $(1/8\sqrt{2})$. Alice thus needs to send at least two sets of her initial preparations for the confirmed transfer of the secret information. Fig. 2 shows the maximal increase in probability amplitude of the desired state after the required number of iteration performed on the initial uniform superposition state.

After performing the operation $-P_\psi$, all the three receivers perform their own local measurements in the computational basis. One of the receivers or at the most two of them may be cheat and may try to get hold of all the three qubits so the honest receiver and the sender must be cautious. However, in the present scheme even if the dishonest receivers succeed in capturing the other two qubits there is a faint possibility of their retrieving the secret information on their own unless they know the correct $-P_{\psi}$ to be performed. Performing a wrong $-U_{\psi}$ will not lead to a substantial increase in amplitude of the desired state.

Furthermore, precise no. of iterations is important. As in the present case, the minimum number of iterations required for increasing the probability of desired state is two. After single iteration the amplitude of the desired state increases to $5/4\sqrt{2}$. If Alice chooses to have only one iteration and she sends the three qubits to the three distant receivers just after performing *phase rotation* operation once, then the chances of getting the desired state by the three receivers is about 78%. Thus to ensure correct retrieval with some degree of confidence Alice needs to send four to five identical sets of qubits. If she goes for a third iteration as well, the probability of the desired state reduces to 33% and thereby increases the probability of rest of the states to 95%. So for again retrieving the desired state we will have to repeat same quantum mechanical steps for a definite number of times. Although there would be some instances where the chances of retrieving the desired state with certainty are maximal but by and large it is observed that with increase in number of qubits the result of measurement is not deterministic so the number of sets of qubits to be initially sent by the sender increases accordingly. Table 2. shows the success probability to reach the desired state after each iteration for two, three, four and five qubits.

**Discussion and concluding remarks**

Grover's search algorithm for an unsorted database containing *N* items retrieves the desired item after performing a sequence of unitary operations on a pure state. Just by having the input in superposition state it has a supremacy over its classical analogue. Its classical counterpart with same amount of hardware would examine each state in the database individually. We can find an object in $O\left(\sqrt{N}\right)$ steps instead of $O(N)$ classical steps. Also if the number of desired states is one-fourth of the total number of elements in the database then the desired state is obtained after single Grover iteration only. This efficaciousness of Grover's algorithm is exploited for the quantum secret sharing protocol. Moreover cheating becomes very much less pronounced in this case. To acquire the whole information alone the eavesdropper or one of the deceptive agents needs all the sets of qubits which in itself is a difficult task. In case he somehow succeeds in acquiring all the sets of qubits, applying the right quantum mechanical inversion about average operator has a very small probability due to the increase in the number of possible states with the number of qubits. And if he applies the wrong inversion about average unitary operation the probability of getting the desired state is considerably reduced, consequently leading to wrong results. Therefore the increased cost due to increase in number of iterations can be compensated by high degree of security.

| $i$ | $\|S_i\rangle$ | $-P_{s_i}\ P_W\|S_1\rangle$ |
|---|---|---|
| 1. | $\|+\rangle\|+\rangle\|+\rangle\|+\rangle$ | $\frac{1}{2}\left(\|0001\rangle+\|0011\rangle+\|0101\rangle+\|0111\rangle\right)$ |
| 2. | $\|+\rangle\|-\rangle\|+\rangle\|-\rangle$ | $-P_W\|S_1\rangle$ |
| 3. | $\|-\rangle\|-\rangle\|+\rangle\|+\rangle$ | $-P_W\|S_1\rangle$ |
| 4. | $\|-\rangle\|-\rangle\|-\rangle\|-\rangle$ | $-P_W\|S_1\rangle$ |
| 5. | $\|+i\rangle\|+i\rangle\|+i\rangle\|+i\rangle$ | $-P_W\|S_1\rangle$ |
| 6. | $\|-i\rangle\|-i\rangle\|-i\rangle\|-i\rangle$ | $-P_W\|S_1\rangle$ |
| 7. | $\|+\rangle\|+\rangle\|+i\rangle\|+i\rangle$ | $\left(\frac{-1}{8}-\frac{i}{8}\right)\left(\|0000\rangle+\|0100\rangle+\|1000\rangle+\|1100\rangle\right)$ $+\left(\frac{3}{8}+\frac{i}{8}\right)\left(\|0001\rangle+\|0101\rangle\right)$ $+\left(\frac{-1}{8}+\frac{i}{8}\right)\left(\|0010\rangle+\|0110\rangle+\|1001\rangle+\|1010\rangle+\|1101\rangle+\|1110\rangle\right)$ $+\left(\frac{1}{8}+\frac{i}{8}\right)\left(\|0011\rangle+\|0111\rangle\right)+\left(\frac{-3}{8}+\frac{i}{8}\right)\left(\|1011\rangle+\|1111\rangle\right)$ |
| 8. | $\|-i\rangle\|+i\rangle\|-i\rangle\|+i\rangle$ | $\left(\frac{-1}{8}+\frac{i}{8}\right)\left(\|0000\rangle+\|0110\rangle+\|1001\rangle+\|1100\rangle+\|1111\rangle\right)$ $+\left(\frac{1}{8}+\frac{i}{8}\right)\left(\|0001\rangle+\|0111\rangle\right)$ $+\left(\frac{-1}{8}-\frac{i}{8}\right)\left(\|0010\rangle+\|1000\rangle+\|1011\rangle+\|1110\rangle\right)$ $+\left(\frac{3}{8}+\frac{i}{8}\right)\|0011\rangle$ $+\left(\frac{-3}{8}+\frac{i}{8}\right)\left(\|0100\rangle+\|1101\rangle\right)$ $+\left(\frac{1}{8}-\frac{i}{8}\right)\|0101\rangle$ $+\left(\frac{-3}{8}-\frac{i}{8}\right)\|1010\rangle$ |
| 9. | $\|-\rangle\|-\rangle\|-i\rangle\|-i\rangle$ | $-P_W\|S_1\rangle$ |
| 10. | $\|+\rangle\|-\rangle\|+i\rangle\|-i\rangle$ | $-P_W\|S_1\rangle$ |

Table 1. The outcomes of some of the $-P_{s_i}\ P_W|S_1\rangle$, where $i$ = *1.....10* represent the possible initial states.

| | 2 qubits | 3 qubits | 4 qubits | 5 qubits |
|---|---|---|---|---|
| 1st iteration | 100% | 78% | 47% | 25% |
| 2nd iteration | | 94.5% | 90% | 60% |
| 3rd iteration | | | 96.1% | 89% |
| 4th iteration | | | | 99.9% |

Table 2. Success probability after each iteration for two, three, four and five qubits.

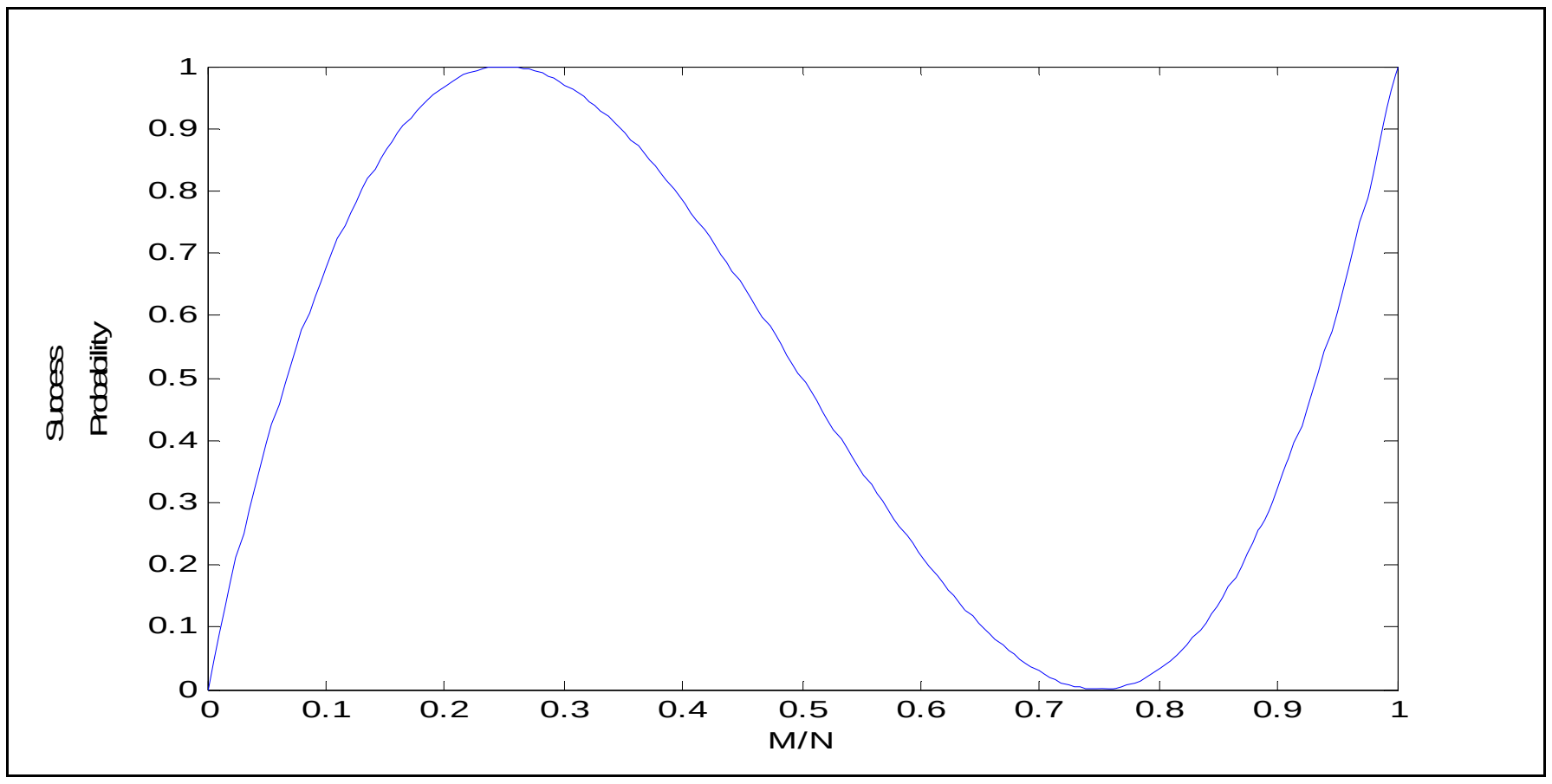

Figure 1: Probability of success of Grover's search algorithm after first iteration.

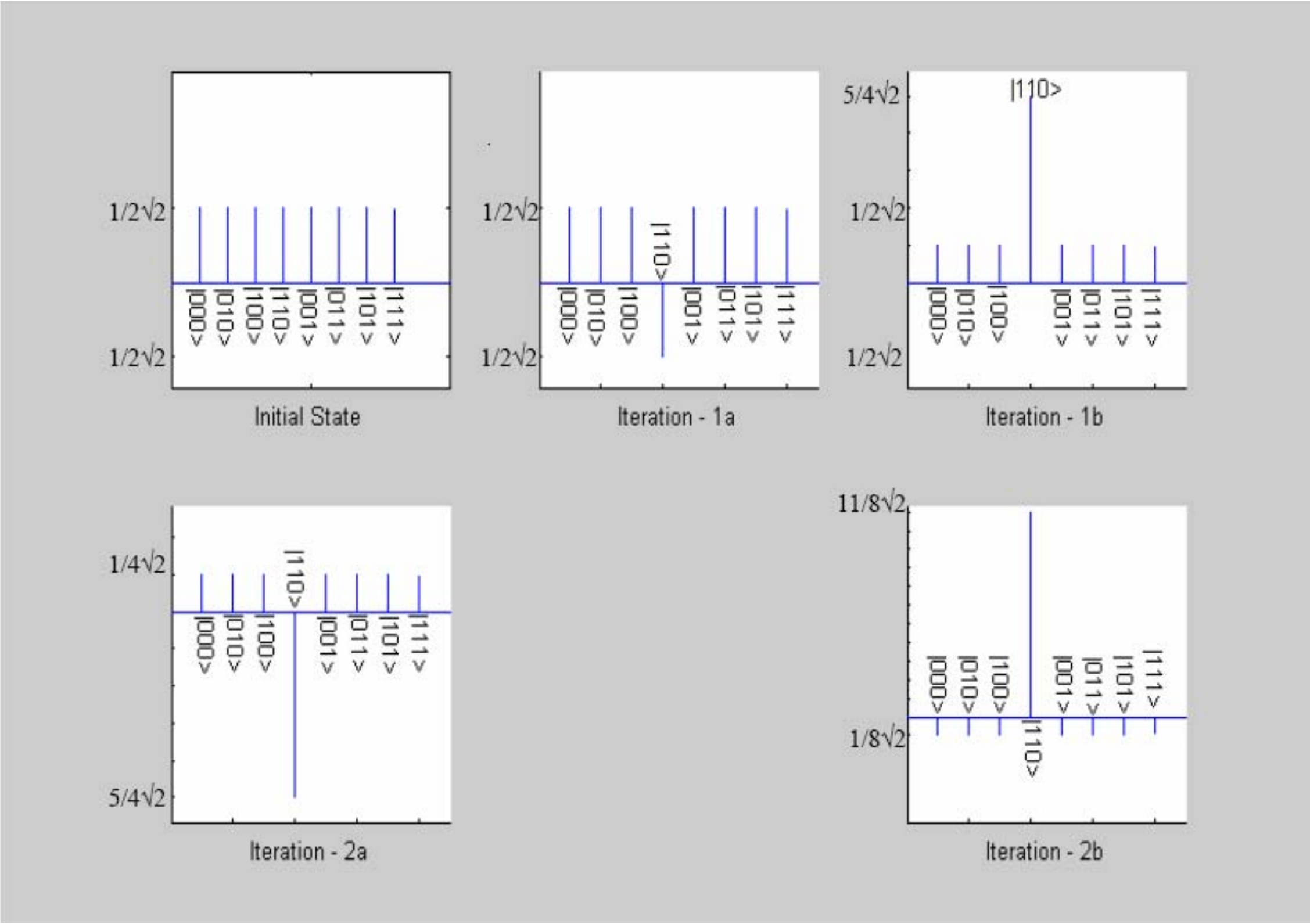

Fig.2 Probability amplitude of the desired state that increases after the required number of iterations performed on the initial uniform superposition state of three qubits.

**Acknowledgement**
Authors are thankful to Dr. M. D. Tiwari for his keen interest and support. Arti Chamoli is thankful is thankful to IIIT, Allahbad for financial support.

**Refrences:**